\documentstyle[aps,psfig]{revtex}

\begin{document}

\preprint{UTF/99-430,ECT*/01-99}
\title{Probing relativistic spin effects in the nucleon by means
of Drell--Yan processes}

\author{F. Cano$^1$, P. Faccioli$^{2,3}$ and  M. Traini$^{1,4}$}

\address{$^1$Dipartimento di Fisica, Universit\`a degli Studi di Trento \\
I-38050 Povo, Italy \\
$^2$ ECT* (European Centre for Theoretical Nuclear Physics
and Related Areas), \\
Villa Tambosi, I-38050 Villazzano (Trento), Italy \\
$^3$ Department of Physics and Astronomy, \\ State University of New York
at Stony Brook, USA. \\
$^4$ Istituto Nazionale di Fisica Nucleare, G.C. Trento.}
\date{\today}
\maketitle
\begin{abstract}
Significant differences between transverse and longitudinal polarized
parton distributions are found at low energies within a light-front
covariant quark model of the nucleon. These differences are due to
relativistic spin effects introduced by the Melosh rotations and
survive evolution to higher $Q^2$ scales. A specific observable
related to double-spin asymmetries in lepton pair production in
polarized hadron-hadron collisions is defined. The possibility of
assessing the relevance of these relativistic spin effects in future
experiments at RHIC and HERA--$\vec{N}$ is discussed.
\end{abstract}


\section{Introduction}
\label{INTRO}

A complete description of the momentum and spin degrees of freedom of
quarks and antiquarks in the nucleon requires, at leading twist, the
definition of three sets of parton distributions.  Two of them, the
momentum distribution $f_1(x,Q^2)$ and the helicity distribution
$g_1(x,Q^2)$, have been intensively investigated in the last few years
(for recent reviews see \cite{LAMPE98,JAFFE96}) while the so called
transversity distribution, $h_1(x,Q^2)$, has come to the attention of
theorists and experimentalists more recently in the analysis of
Drell--Yan spin asymmetries \cite{RALSTON79}.  The main reason why it
has passed unnoticed for such a longtime is related to its chiral-odd
character.  If quark masses are neglected in the QCD Lagrangian, no
interaction at lowest order of perturbation theory, can flip
chirality.  As a consequence, transversity is strongly suppressed (by
powers of $m_q/Q$) in deep inelastic lepton-nucleon scattering (DIS)
and in general in any hard process that involves only one parton
distribution. In hadron-hadron collisions the chirality of the partons
that annihilate is uncorrelated and the previous restrictions do not
apply.

Recently there has been a number of proposals to measure the
transversity parton distributions (see \cite{JAFFE97} for a
review). Lepton pair production in doubly polarized Drell-Yan
processes are among the proposed scenarios and such experiments are
included in the research program at RHIC and the HERA-$\vec{N}$ project
\cite{SAITO98}.  The feasibility of measurements of double transverse
asymmetries ($A_{TT}$) at polarized $pp$ colliders (RHIC) and
fixed-target experiments (HERA-$\vec{N}$) has been recently studied
\cite{MARTIN98}. The expected maximal value for $A_{TT}$ at the
kinematic range covered by RHIC is around 1-2 \%, which would be
difficult to measure with the present acceptances.  For
HERA-$\vec{N}$, $A_{TT}$ is expected to reach values as large as 5 \%.

Along with the experimental prospects, many efforts have also been
made on the theoretical side
\cite{JAFFE91,MA98,SCHMIDT97,SCOPETTA98,BARONE97,MODELS}.  In
particular Jaffe and Ji calculated $h_1$ within the Bag Model of the
nucleon \cite{JAFFE91}, pointing out the relativistic character of the
differences between transverse and longitudinal polarization
properties.

The calculation of leading order (LO) \cite{ARTRU90} and
next-to-leading order (NLO) \cite{VOGELSANG98} anomalous dimensions
allowed to address quantitatively the differences between $g_1$ and
$h_1$ due to perturbative QCD (pQCD) evolution in different models
\cite{SCOPETTA98,BARONE97}.  In particular, Scopetta and Vento have
shown that pQCD differences are sizeable only for low values of $x$
($x \lesssim 0.1$) \cite{SCOPETTA98}.

On the other hand the boundary conditions of the evolution equations,
that are provided by the low energy input at the hadronic scale
$Q_0^2$ ($\lesssim 1$ GeV$^2$), can yield an additional, non
perturbative, difference.  As it was stressed by Jaffe and Ji
\cite{JAFFE91}, at this scale the equality $h_1(x,Q_0^2) =
g_1(x,Q_0^2)$ is a typical outcome of non-relativistic models of the
nucleon, in which motion and spin observables are uncorrelated.  In
other words, any departure from the previous identity is a signature
of relativity in the employed hadronic model. A complete theoretical
study of $h_1$ and $g_1$ has to account for both: the relativistic
effects which distinguish $h_1$ from $g_1$ at the non-perturbative
scale, and the pQCD evolution which differs for the two structure
functions.

	Aim of the present work is the quantitative study of the
relativistic effects in $h_1$ and $g_1$ due to the correlations of
spin and parton motion in the hadronic systems. For this purpose we
use light-front dynamics formalism in which the interplay between
motion and spin is made explicit through the Melosh rotations. In
particular, we make use of the light-front covariant (LFC) quark model
of ref. \cite{FACCIOLI98} to compute the leading twist contribution to
the matrix elements at the hadronic scale $Q^2_0$. The
non-perturbative input is then evolved, at NLO, up to a higher $Q^2$
scale. We shall show that relativistic corrections introduced at
$Q^2_0$ clearly survive evolution. Measurements of observables
involving transverse and longitudinal asymmetries might put in
evidence the corrections to the naive non-relativistic spin picture
adopted in low-energy models of the nucleon.

The paper is organized as follows: in section \ref{LFC} the framework
to calculate $h_1$ and $g_1$ within the LFC quark model is described
and a specific observable, related to double spin asymmetry
experiments, is defined. The definition is such that the view on
relativistic spin effects is optimized. Results in the kinematic range
covered by RHIC and HERA-$\vec{N}$ are presented in section \ref{RES},
where the feasibility of the detection of these effects is also
discussed.  Finally, conclusions are drawn in section \ref{CONCL}.

\section{Polarized parton distributions in a LFC quark model}
\label{LFC}

\subsection{Polarized partons at the hadronic scale}
\label{polpartons}

        The helicity distribution $g_1^a(x,Q^2)$ of a parton with
flavour $a$ is defined as the probability of having a (longitudinally)
polarized parton $a$ with spin parallel to the longitudinal
polarization of the parent nucleon minus the probability of finding
the parton polarized in the opposite direction. A similar definition
applies to $h_1^a(x,Q^2)$ for transverse polarizations \cite{JAFFE96}.

        In the quark model a simple approach can be developed to
connect parton and momentum densities at the hadronic scale $Q_0^2$
where valence degrees of freedom dominate the matrix elements at
leading twist \cite{TRAINI97}. In particular, $g_1^a$
\cite{FACCIOLI98,TRAINI97} and $h_1^a$ \cite{SCOPETTA98} can be
related to the longitudinal and transverse quark momentum densities
respectively. Namely:

\begin{eqnarray}
g_1^a(x,Q^2_0) & = & \frac{1}{(1-x)^2} \int d^3 k \; (n_a ^\uparrow
(\vec{k}) - n_a ^\downarrow (\vec{k})) \delta \left( \frac{x}{1-x} -
\frac{k^+ }{M/\sqrt{2}}\right)
\label{g1def} \; , \\
h_1^a(x,Q^2_0) & = & \frac{1}{(1-x)^2} \int d^3 k \; (n_a ^\rightarrow
(\vec{k}) - n_a ^\leftarrow (\vec{k})) \delta \left( \frac{x}{1-x} -
\frac{k^+ }{M/\sqrt{2}}\right) \; ,
\label{h1def}
\end{eqnarray}

\noindent where $\vec{k}$ is the three-momentum of the struck quark,
$M$ is the nucleon mass, $k^+=(\sqrt{\vec{k}^2 + m_q^2} +
k_z)/\sqrt{2}$ and $n_a^{\uparrow (\downarrow)}(\vec{k})$ is the
density of (valence) quarks with momentum $\vec{k}$ and longitudinal
polarization aligned (antialigned) with the longitudinal polarization
of the parent nucleon. A similar notation
($n_a^{\stackrel{\longrightarrow}{\leftarrow}}(\vec{k})$) is used for
transverse polarization. The explicit expressions for the densities
are:

\begin{eqnarray}
n_a^{\uparrow \downarrow}(\vec{k}) & = & \langle P, S_z=+1/2 |
\sum_{i=1}^3 {\cal{P}}_i^a \frac{1 \pm \sigma_z^{(i)}}{2} \delta
(\vec{k} - \vec{k_i}) | P, S_z=+1/2 \rangle \; ,
\label{naupdef} \\
n_a^{\stackrel{\longrightarrow}{\leftarrow}}(\vec{k}) & = & \langle P,
S_x=+1/2 | \sum_{i=1}^3 {\cal{P}}_i^a \frac{1 \pm \sigma_x^{(i)}}{2}
\delta (\vec{k} - \vec{k_i}) | P, S_x=+1/2 \rangle \; ,
\label{nafwdef}
\end{eqnarray}

\noindent where ${\cal{P}}_i^a$ is the flavour projector.

        In any non-relativistic description of the wave function the
densities (\ref{naupdef}) and (\ref{nafwdef}) are equal and hence
$h_1(x,Q^2_0) = g_1(x,Q^2_0)$ as discussed by Jaffe and Ji
\cite{JAFFE91}. However the degeneracy is removed by relativistic
covariance requirements, such as those which are implemented in the
covariant quark models based on light-front dynamics
\cite{FACCIOLI98}.

Light-front dynamics with a fixed number of particles has been widely
discussed in the literature (see \cite{KEISTER91} for reviews), so
that it is enough to highlight briefly the main features of the
angular momentum in this formalism.  In relativistic quantum theories,
the angular momentum operator is obtained from the generators of the
Poincar\'e group.  In the light front form of the dynamics problems
arise when considering addition of angular momenta, because usual
composition rules are not satisfied.  Nevertheless, it is possible to
restore them by means of a unitary transformation: the Melosh rotation
\cite{MELOSH74}. From the physical viewpoint these transformations
relate the angular momentum eigenstates in a subsystem rest frame (a
quark, for example) to the centre of mass frame.  For spin 1/2
particles the Melosh rotations (MR) that link light-front to canonical
spin states are represented by:

\begin{equation}
D^{1/2}[R_M(\vec{k})] = \frac{(m+\omega+k_z) - i \vec{\sigma} \cdot
(\hat{z} \times \vec{k}_\perp)}{( (m+\omega+k_z)^2 +
\vec{k}_\perp^{\,2})^{1/2}} \;\; ,
\label{mr}
\end{equation}

\noindent where $\vec{k}$ is the three-momentum of the particle and
$\omega = \sqrt{\vec{k}^{\,2} + m^2}$ its relativistic energy.  The
fact that $D^{1/2}[R_M(\vec{k})] \rightarrow 1$ in the limit
$\vec{k}_\perp \rightarrow 0$ reveals the relativistic origin of the
MR. A new spin-flavor basis is then defined (the Pauli-Melosh basis)
in which the ordinary (canonical) Pauli spinors for each individual
particle are replaced by the Melosh-rotated spinors:

\begin{equation}
\chi_i = D^{1/2}[R_M(\vec{k}_i)] \chi_i^c =
\frac{(m_i+\omega_i+k_{i_z}) - i \vec{\sigma}^{(i)} \cdot (\hat{z}
\times \vec{k}_{i_\perp})}{( (m_i+\omega_i+k_{i_z})^2 +
\vec{k}_{i_\perp}^{\,2})^{1/2}} \chi_i^c \; ,
\label{melosh2}
\end{equation}

\noindent where $\chi_i^c$ are the usual (canonical) Pauli spinors for
the particle $i$. The Pauli-Melosh basis constitutes a 'minimal'
relativistic spin basis \cite{BEYER98} that, despite of its
non-manifest covariance, is compatible with Poincar\'e invariance
requirements.

In our calculation we shall refer to the wave function $\Psi$ of ref.
\cite{FACCIOLI98}, where the relativistic mass equation:

\begin{equation}
(M_0 + V) \; \Psi \equiv (\sum_{i=1}^3 \sqrt{\vec{k}_i^{\, 2} + m_i^2}
+ V) \; \Psi = E \Psi
\label{rme}
\end{equation}

\noindent has been solved. The free relativistic mass operator
$M_0=\sum_{i=1}^3 \omega_i$ is supplemented with an hypercentral
phenomenological interaction

\begin{equation}
V = - \frac{\tau}{\xi} + \kappa_l \xi +\Delta \; ,
\label{potential}
\end{equation}

\noindent 
where the hyper-radius $\xi^2 = \sum_{i=1}^3 (\vec{r}_i - \vec{R})^2$
is a function of what, in a non relativistic treatment, is the vector
position of the particles $\vec{r}_i$ and the center of mass position
$\vec{R}$. $\tau$, $\kappa_l$ and $\Delta$ are constants fixed to
reproduce the basic features of the low-energy baryonic spectrum
pattern in the $J^P=1/2^\pm$ channels.  Since this interaction is
invariant under rotations and does not depend on the total light-front
momentum, all the commutation relations that guarantee the covariance
requirement (see \cite{KEISTER91}) are correctly satisfied. The
advantage of the hypercentral interactions, well known in the
non-relativistic limit \cite{FERRARIS95}, is that they allow a
straightforward solution of the Schr\"odinger equation. The resulting
spin-isospin wave function is $SU(6)$ symmetric. It should be
emphasized that the degeneracy between $h_1$ and $g_1$ in the
non-relativistic limit is independent of the presence of
$SU(6)$--breaking terms in the nucleon wave function.

        From eq. (\ref{melosh2}) it is evident the correlation between
spin and motion in the Pauli-Melosh spin basis, so that the momentum
densities (\ref{naupdef}) and (\ref{nafwdef}) are no longer equal,
independently of its SU(6)-symmetric character.  Differences between
$h_1^a(x,Q^2_0)$ and $g_1^a(x,Q^2_0)$ are now proportional to the
internal transverse momentum $\vec{k}_\perp$, as expected:

\begin{eqnarray}
h_1^u(x,Q_0^2) - g_1^u(x,Q_0^2) & = & - 4 (h_1^d(x,Q_0^2) -
g_1^d(x,Q_0^2)) \nonumber \\ & = & \frac{1}{(1-x)^2} \int d^3 k \;
\left[ \frac{4}{9} \frac{\vec{k}_\perp^{\, 2}}{(m+\omega+k_z)^2 +
\vec{k}_\perp^{\,2})} \right] \nonumber \\ & & \times n(\vec{k} )\;
\delta \left( \frac{x}{1-x} - \frac{k^+ }{M/\sqrt{2}}\right) \; ,
\label{diff}
\end{eqnarray}

\noindent where $n(\vec{k})$ is the unpolarized flavorless quark
momentum density normalized to the number of particles as obtained by
solving the mass equation (\ref{rme}). Equation (\ref{diff}) is one of
the central results of the present work. It shows that differences
between $h_1^a$ and $g_1^a$ have a clear relativistic origin depending
on the kinematical structure of the MR. The $x$-dependence of the
quantity $h_1^a - g_1^a$ is embodied in the wave function term
$n(\vec{k})$. Within a fully non-relativistic approach $h_1^a - g_1^a$
would vanish and the actual (and identical) values of $h_1^a$ and
$g_1^a$ would be largely suppressed for $x \gtrsim 0.5$ due to the
lack of high momentum components in $n(\vec{k})$ for typical
non-relativistic models \cite{ROPELE...}. In fact the relativistic
effects show up, in the present approach, in a twofold way: i) MR
introduce the non-vanishing difference $h_1^a - g_1^a$ of
eq.(\ref{diff}); ii) high momentum components in $n(\vec{k})$ which
are mostly originated by the relativistic kinetic energy in the mass
equation (\ref{rme}) and that reinforce the MR contribution, as
discussed in ref.\cite{FACCIOLI98} for $g_1^a$.

        Concerning the scale dependence, it should be stressed that
eq.(\ref{diff}) is valid at a low-energy (hadronic) scale $Q_0^2$,
where the non-perturbative models used to calculate the twist-two
matrix elements can be applied. At higher (experimental) scales
$Q^2>Q_0^2$ perturbative QCD evolution yields an additional
contribution to the differences between longitudinal and transverse
polarization observables (cfr. section \ref{RES}).  Previous attempts
\cite{MA98} to study the effects of MR in $h_1$ disregarded this scale
dependence and did not include perturbative contributions.

\subsection{Double Spin Asymmetries}
\label{DSA}

Let us now discuss a specific combination of longitudinally and
transversely polarized parton distributions, namely the ratio

\begin{equation}
R_{TL}(x_1,x_2,Q^2) = \frac{\sum_a e_a^2 h_1^a(x_1,Q^2)
h_1^{\bar{a}}(x_2,Q^2) + (x_1 \leftrightarrow x_2)} {\sum_a e_a^2
g_1^a(x_1,Q^2) g_1^{\bar{a}}(x_2,Q^2) + (x_1 \leftrightarrow x_2)} \;
,
\label{rdef}
\end{equation}

\noindent
where $e_a$ is the charge of a quark of flavour $a$.  The arguments
$x_1$ and $x_2$ are related, for Drell-Yan processes, to the center of
mass energy $\sqrt{s}$, the invariant mass of the produced lepton pair
$Q^2$, and the rapidity $y=\arctan(Q^3/Q^0)$:

\begin{eqnarray}
x_1 = \sqrt{\frac{Q^2}{s}} e^y & \;\;\;\; & x_2 = \sqrt{\frac{Q^2}{s}}
e^{-y} \; \; .
\label{xdefs}
\end{eqnarray}

$R_{TL}(x_1,x_2,Q^2)$ does not dependent on the unpolarized parton
distribution and involves only the ratio between $g_1$ and $h_1$. To
this respect it is less model dependent than the single terms and is
quite transparent for studying MR effects.

 The flavor combination of transverse parton distributions in
Eq. (\ref{rdef}) appears in the LO double transverse asymmetry of
Drell-Yan processes:

\begin{eqnarray} 
\left. A_{TT}\right|_{LO} & = &
\left. \frac{\frac{d\sigma(\stackrel{\longrightarrow}{\rightarrow})}{dQ^2
\, dy\, d\phi} -
\frac{d\sigma(\stackrel{\longleftarrow}{\rightarrow})}{dQ^2 \, dy\,
d\phi}} {\frac{d\sigma(\stackrel{\longrightarrow}{\rightarrow})}{dQ^2
\, dy} + \frac{d\sigma(\stackrel{\longleftarrow}{\rightarrow})}{dQ^2
\, dy}}\right|_{LO} = \nonumber \\ & = & \frac{\cos 2 \phi }{4 \pi}
\left( \frac{\sum_a e_a^2 h_1^a(x_1,Q^2) h_1^{\bar{a}}(x_2,Q^2) + (x_1
\leftrightarrow x_2)} {\sum_a e_a^2 f_1^a(x_1,Q^2)
f_1^{\bar{a}}(x_2,Q^2) + (x_1 \leftrightarrow x_2)} \right),
\label{attdef}
\end{eqnarray}

\noindent 
where the arrows denote the transverse polarization of the beam and
target and $\theta$ ($\phi$) is the polar (azimuthal) angle of the
detected lepton.

    The corresponding LO asymmetry with longitudinally polarized
hadrons, $A_{LL}$, is proportional to the combination of helicity
distributions $g_1$ considered in Eq. (\ref{rdef}):

\begin{equation}
\left. A_{LL}\right|_{LO} = \left.\frac{ \frac{d\sigma(\uparrow
\downarrow)}{dQ^2 \, dy} - \frac{d\sigma(\uparrow \uparrow)}{dQ^2 \,
dy}} {\frac{d\sigma( \uparrow \uparrow)}{dQ^2 \, dy} +
\frac{d\sigma(\uparrow \downarrow)}{dQ^2 \, dy}}\right|_{LO} =
\frac{\sum_a e_a^2 g_1^a(x_1,Q^2) g_1^{\bar{a}}(x_2,Q^2)+ (x_1
\leftrightarrow x_2)} {\sum_a e_a^2 f_1^a(x_1,Q^2)
f_1^{\bar{a}}(x_2,Q^2) + (x_1 \leftrightarrow x_2)}\; ,
\label{alldef}
\end{equation}

	Then, $R_{TL}$ is just the ratio between double transverse and
	double longitudinal asymmetries at LO:

\begin{equation}
R_{TL} = \frac{4 \pi}{ \cos 2 \phi} \frac{A_{TT}|_{LO}}{A_{LL}|_{LO}}
\end{equation}

The evolution procedure we are going to use is a NLO procedure and
therefore the parton distribution and the corresponding cross sections
should be consistently evaluated at NLO. As a consequence the
asymmetries do not assume the simple forms (\ref{attdef}) and
(\ref{alldef}). Nevertheless it is still feasible to establish a
rather close connection between the combinations that enter 
Eq. (\ref{rdef}) and the NLO asymmetries.

        Indeed, the differential cross section for longitudinally
polarized process is written, at NLO, as \cite{GEHRMANN97}:

\begin{eqnarray}
\left. \frac{d\sigma(\uparrow \downarrow)}{dQ^2 \, dy} -
\frac{d\sigma(\uparrow \uparrow)}{dQ^2 \, dy}\right|_{NLO} & = &
\frac{4 \pi \alpha_{\mbox {\tiny em}}^2}{9 s Q^2} \int dx_1 \; dx_2 \;
\left\{\phantom{.}  \right. \nonumber \\ & & \left(\sum_a e_a^2
g_1^a(x_1,Q^2) g_1^{\bar{a}}(x_2,Q^2)\right)\, \left(\Delta
c_{q\bar{q}}^{\mbox {\tiny DY} (0)} + \frac{\alpha_s(Q^2)}{2 \pi}
\Delta c_{q\bar{q}}^{\mbox {\tiny DY} (1)}\right) \nonumber \\ & + &
\Delta G(x_1,Q^2) \sum_a e_a^2 \left(g_1^a(x_2,Q^2) +
g_1^{\bar{a}}(x_2,Q^2)\right) \frac{\alpha_s(Q^2)}{2 \pi} \Delta
c_{qg}^{\mbox {\tiny DY} (1)} \nonumber \\ & + & \left. (x_1
\leftrightarrow x_2) \right\} \; ,
\end{eqnarray}

\noindent
where $\Delta G(x,Q^2)$ is the polarization of gluons and the
coefficient $\Delta c_{q\bar{q}}^{\mbox {\tiny DY} (0)} =
\delta(x_1-\sqrt{\frac{Q^2}{s}} e^y) \delta(x_2-\sqrt{\frac{Q^2}{s}}
e^{-y})$. The NLO order coefficients are explicitly given in
\cite{GEHRMANN97}. The NLO order introduces on the one hand the
polarization of the gluons due to the subprocess $qg \rightarrow
q\gamma$ and on the other hand the coefficients $\Delta c_{q\bar{q},
qg}^{\mbox {\tiny DY} (1)}$ that contain terms which are not
proportional to $\delta(x_1-\sqrt{\frac{Q^2}{s}} e^y)
\delta(x_2-\sqrt{\frac{Q^2}{s}} e^{-y})$ and therefore break the
relationship (\ref{xdefs}) we have used in our calculation. A similar
modification appears for the unpolarized DY cross
section. Nevertheless, an estimation of the asymmetry $A_{LL}$ by
using available parameterization of polarized parton distributions
indicate that the NLO asymmetry is dominated by the $O(\alpha_s^0)$ of
the $q\bar{q} \rightarrow \gamma$ subprocess in the central rapidity
region. This term contains just the combination of polarized parton
distributions that we are interested in and which enters in the ratio
(\ref{rdef}). Alternatively, it would also be possible to extract this
combination of helicity parton distributions from other experiments.

        In the transversely NLO polarized cross sections, gluons are
absent since there is no equivalent 'transversity' for gluons, and
hence \cite{VOGELSANG93}:

\begin{eqnarray}
\left.\frac{d\sigma(\stackrel{\longrightarrow}{\rightarrow})}{dQ^2 \,
dy\, d\phi} -
\frac{d\sigma(\stackrel{\longleftarrow}{\rightarrow})}{dQ^2 \, dy\,
d\phi}\right|_{NLO} & = & \frac{\alpha_{\mbox {\tiny em}}^2}{9 s Q^2}
\cos(2 \phi) \int dx_1 \; dx_2 \; \left\{\phantom{.} \right. \nonumber
\\ & & \left(\sum_a e_a^2 h_1^a(x_1,Q^2)
h_1^{\bar{a}}(x_2,Q^2)\right)\, \left(\delta c_{q\bar{q}}^{\mbox
{\tiny DY} (0)} + \frac{\alpha_s(Q^2)}{2 \pi} \delta
c_{q\bar{q}}^{\mbox {\tiny DY} (1)}\right) \nonumber \\ & + &
\left. (x_1 \leftrightarrow x_2) \right\} \;\;\; .
\end{eqnarray}

        Though the cross section is proportional to the combination of
transversity distributions that we have used in (\ref{rdef}), the NLO
order coefficient $\delta c_{q\bar{q}}^{\mbox {\tiny DY} (1)}$ may
spoil the relationship (\ref{xdefs}) again. However, it was shown in
ref.\cite{VOGELSANG93}, that the dominant term in this coefficient
(and in the equivalent one for the unpolarized cross section) is
proportional to $\delta(x_1-\sqrt{\frac{Q^2}{s}} e^y)
\delta(x_2-\sqrt{\frac{Q^2}{s}} e^{-y})$, so that it is still possible
a rather direct extraction of the combination $\sum_a e_a^2
h_1^a(x_1,Q^2) h_1^{\bar{a}}(x_2,Q^2)$ from the analysis of the
transverse asymmetry at NLO.

\section{Results}
\label{RES}

\subsection{Parton distributions: non-perturbative and perturbative
contributions}

\label{PDFQ0}

        To reach the experimental regime we have evolved the
calculated parton distributions from the hadronic scale up to a scale
$Q^2 > Q_0^2$ by using pQCD at NLO, in the $\overline{\mbox{MS}}$
scheme.  The sensitivity to the factorization scheme in this approach
has been tested in previous calculations \cite{FACCIOLI98} and found
to be generally small.

The value of $Q_0^2$ (the hadronic scale) is fixed \cite{TRAINI97} by
evolving backwards the parametrized experimental fits for unpolarized
parton distributions \cite{CTEQ4}, to the scale where the valence
quarks carry the whole momentum of the nucleon. We found $Q_0^2=0.094$
GeV$^2$ and the reliability of the pQCD evolution procedure at such
low scale was studied in detail in
\cite{SCOPETTA98,FACCIOLI98,TRAINI97}.

Let us emphasize that the value $Q_0^2 = 0.094$ GeV$^2$ is found by
using an evolution code where the transcendental equation

\begin{equation}
\ln {Q^2 \over \Lambda_{NLO}^2} - {4\pi \over \beta_0
\alpha_s(Q^2)} + {\beta_1 \over \beta_0^2} \ln \left[{4\pi \over
\beta_0 \alpha_s(Q^2)} + {\beta_1 \over \beta_0^2} \right] =0
\label{trasceq}
\end{equation}

\noindent 
has been solved to get the NLO coupling constant.  The use of the full
NLO expression (\ref{trasceq}) is mandatory when evolving from/to a
low energy scale \cite{WEIGL96} and the approximate solution

\begin{equation}
\alpha_s(Q^2) = {1 \over \beta_0 \ln (Q^2/\Lambda_{NLO}^2)}
\left[1 - {\beta_1 \over \beta_0^2}{\ln\ln(Q^2/\Lambda_{NLO}^2)
\over \ln(Q^2/\Lambda_{NLO}^2)}\right] \;\; ,
\label{appeq}
\end{equation}

\noindent which is valid for large values of $Q^2/\Lambda_{ NLO}^2$
and often used in NLO codes (e.g. ref.\cite{HIRAI98}), would introduce
quite spurious effects. In particular, the evolution to energy scales
as low as the hadronic point $Q_0^2$ may yield a negative gluon
contribution at $Q_0^2$, related to a large extent to the spurious
effects introduced by eq.(\ref{appeq}). In the present case, for
example, evolving back the experimental fit \cite{CTEQ4} to the
hadronic scale $Q_0^2 = 0.094$ GeV$^2$, would give a fraction of
momentum carried by the gluons of $-1.44 \times 10^{-2}$, which,
although non-zero, is relatively small. Obviously, the
situation is much worse when Eq. (\ref{appeq}) is used instead of
Eq. (\ref{trasceq}).

In addition, the evolution code to be used for hadronic model
calculations must guarantee complete symmetry for the forward and
backward paths $Q_0^2 \to Q^2$ and vice versa, as implied by genuine
perturbative QCD expansion at NLO. Additional approximations
associated with Taylor expansion valid for large $Q^2/\Lambda_{NLO}^2$,
 must be avoided, as discussed in \cite{TRAINI97}.

In Fig. 1 we show the curves for $h_1^u$ and $g_1^u$ at the hadronic
scale $Q_0^2$ (Fig. 1(a)) and at the partonic scale $Q^2=100$ GeV$^2$
(Fig. 1(b)). A remarkable difference between $x h_1(x,Q_0^2)$ and $x
g_1(x,Q_0^2)$ appears at large $x$, reaching a peak at $x \approx
0.5$. Quantitatively they are bigger that those obtained within bag
models \cite{JAFFE91,SCOPETTA98}. It is clear that the probability of
transverse polarization is larger than the longitudinal one when
relativistic effects are considered.

It is possible to appreciate better the MR effects by simply setting
$D^{1/2}[R_M(\vec{k})] \rightarrow 1$. In this case the remaining
relativistic ingredient is the high momentum components generated by
the 'relativized' Schr\"odinger equation (\ref{rme}). The results are
also shown in Fig. 1(a) and we get $h_1(x,Q_0^2)=g_1(x,Q_0^2)$ as
expected. After evolution (Fig. 1(b)) we see that when the MR are
neglected, $g_1$ and $h_1$ differ mainly at low $x$ ($x \lesssim 0.1$)
because of pQCD evolution \cite{SCOPETTA98,BARONE97}, while the
inclusion of the correlations between spin and motion produce large
effects also in the medium and large $x$ region.

In order to study the effects of MR in the combination of polarized
parton distributions we are interested in, let us first consider the
ratios

\begin{equation}
R_L=\frac{\sum_a e_a^2 g_1^a(x_1,Q^2) g_1^{\bar{a}}(x_2,Q^2) + (x_1
\leftrightarrow x_2)} {\sum_a e_a^2 f_1^a(x_1,Q^2)
f_1^{\bar{a}}(x_2,Q^2) + (x_1 \leftrightarrow x_2)}
\label{rldefinition}
\end{equation}

\noindent and

\begin{equation}
R_T = \frac{\sum_a e_a^2 h_1^a(x_1,Q^2) h_1^{\bar{a}}(x_2,Q^2) + (x_1
\leftrightarrow x_2)} {\sum_a e_a^2 f_1^a(x_1,Q^2)
f_1^{\bar{a}}(x_2,Q^2) + (x_1 \leftrightarrow x_2)} \;\;\; ,
\label{rtdefinition}
\end{equation}

\noindent shown in Fig. 2(a) and Fig. 2(b) respectively as a function
of the center of mass rapidity $y$ in a kinematic region ($\sqrt{s} =
100$ GeV, $Q^2=100$ GeV$^2$) accessible by RHIC experiments. These
values are of the order of a few percent, compatible with those
obtained in the literature (see for instance \cite{BARONE97}, where
antiquarks are considered also at the hadronic scale). Again it is
possible to single out the contribution introduced by MR switching
them off. In general, MR reduce both $R_{T}$ and $R_{L}$ but this
reduction is far more significant for $R_{L}$.

  In Fig. 3 we show the ratio $R_{TL}=R_T/R_L$ as a function of $y$
(in the same kinematical region) when considering Melosh rotations and
when they are neglected. Figure 3 represents our reference point to
assess the importance of relativistic spin effects: they enhance the
transverse ratio $R_T$ with respect to the longitudinal term $R_L$.
Namely, if Melosh rotations are taken into account $R_{T} \simeq
R_{L}$ in the considered kinematic regime.  On the contrary, if the
spin-motion correlations are neglected (non-relativistic limit), then
in the same region $R_{T} \simeq \frac{1}{2} R_{L}$.  Experiments may
decide between these two alternatives, therefore probing the relevance
of covariance effects in the dynamical description of the nucleon
spin.

	Before examining the possibility of measuring these effects
	let us discuss some issues concerning the previous results:
	the coupling of $g_1$ to the polarized gluons in the evolution
	at NLO and the dependence of $R_{TL}$ on the spatial nucleon
	wave function.

\subsection{LO versus NLO evolution}

        In our valence model of the nucleon, the contribution to
$g_1^{\bar{q}}(x,Q^2)$ and $h_1^{\bar{q}}(x,Q^2)$ comes from evolution
only. On qualitative grounds one would expect $h_1^{\bar{q}}(x,Q^2) <<
g_1^{\bar{q}}(x,Q^2)$ (i.e. $R_{TL} << 1$), since $h_1^{\bar{q}}$ at
LO vanishes while $g_1^{\bar{q}}$ receives contributions also from the
lowest order in $\alpha_s$.  However, a careful analysis shows that
this is not necessarily the case, since the final value of
$g_1^{\bar{q}}(x,Q^2)$ is not only determined by the order of
$\alpha_s$ but also from the behavior of the anomalous dimensions.
More specifically, in order to understand why one gets
$g_1^{\bar{q}}(x,Q^2) \approx h_1^{\bar{q}}(x,Q^2)$, let us consider
the DGLAP equations for the longitudinally polarized parton
distributions \cite{GLUCK96} to disentangle the coupling of
$g_1^{\bar{q}}$ to quarks, antiquarks and gluons\footnote{This kind of
analysis is far more cumbersome in the evolution procedure based on
the RGE.}:

\begin{equation}
\frac{\partial}{\partial\ln Q^2} \, \Delta q_{_{NS}}(x,Q^2)\ =
\frac{\alpha_s (Q^2)}{2\pi} \ \Delta P_{q^\pm, {NS}} (x) \otimes
\Delta q_{_{NS}}(x,Q^2)
\end{equation}

\begin{eqnarray}
\frac{\partial}{\partial \ln Q^2} \left(\begin{array}{c} \Delta {q}_s
(x,Q^2) \\ \Delta g (x,Q^2)
\end{array} \right) & = & \frac{\alpha_s (Q^2)}{2\pi}
\left( \begin{array}{cc} \Delta P_{qq}(x,Q^2) & \Delta P_{qg}(x,Q^2)
  \\ \Delta P_{gq}(x,Q^2) & \Delta P_{gg}(x,Q^2) \\
\end{array} \right) \otimes
\left( \begin{array}{c} \Delta {q}_s (x,Q^2) \\ \Delta g(x,Q^2)
\end{array} \right)
\end{eqnarray}

\noindent where $\Delta {q}_s=\sum_a(g_1^a +g_1^{\bar{a}})$ and
$\Delta q_{_{NS}}$ corresponds to non-singlet combinations of $g_1^a$
and $g_1^{\bar{a}}$.  For the sake of simplicity we will give results
for $\frac{\partial g_1^{\bar{u}}(x,Q^2)}{\partial \ln Q^2}$ at
$Q^2=100$ GeV$^2$ (conclusions are not changed for other $Q^2$ points
we have checked) at LO.

        In Fig. 4 we show the contribution to the derivative of
$g_1^{\bar{u}}$ due to the coupling to $\Delta G(x,Q^2)$ and to the
terms proportional to the quarks and antiquarks at LO. The remarkable
feature is the opposite sign of the coupling to the gluons and to the
quarks+antiquarks at LO that makes the final derivative be not too
large for $x \gtrsim 0.05$. This partial cancellation between
different contributions produces a rather slow increase of
$g_1^{\bar{u}}(x,Q^2)$ with $Q^2$ at order $(\alpha_s/2\pi)$ and the
final value of $g_1^{\bar{u}}(x,Q^2)$ at NLO are comparable to that of
$h_1^{\bar{u}}(x,Q^2)$. This happens for $x \gtrsim 0.05$, while for
smaller $x$ the growing of $g_1^{\bar{u}}$ dominates over that of
$h_1^{\bar{u}}$, due to some extent to the large size of the coupling
to gluons, which is absent in the evolution of $h_1^{\bar{u}}$.

	It should be reminded that $R_{TL}$ vanishes at leading order
since the sea quarks are entirely generated through evolution. This
fact prevent us from doing a quantitative comparison between the LO
and NLO for $R_{TL}$. However, Scopetta and Vento \cite{SCOPETTA98}
analyzed the LO and NLO evolution for other flavour combinations of
$h_1$ and $g_1$ and it was found that these differences between LO and
NLO were not too large and went in the same direction for $g_1$ and
$h_1$.

\subsection{Quark model dependence}

        At this point one can argue that these results for
$R_{TL}(x_1,x_2,Q^2)$ could be strongly influenced by the presence of
high momentum components in the spatial part of the wave function,
carried over by the presence of a relativistic kinetic energy in the
mass operator.  These components have been proved to be essential to
describe electromagnetic transitions and elastic form factors
\cite{CARDARELLI95} and also the high $x$ region in DIS
\cite{FACCIOLI98,TRAINI97}. We have therefore repeated the calculation
of the ratio $R_{TL}(x_1,x_2,Q^2)$ starting from a nucleon wave
function obtained by making use of a a non-relativistic kinetic energy
in the mass operator (cfr. eq. (\ref{rme})).  The interaction has been
kept of the same form (\ref{potential}), while the values of the
parameters ($\tau$, $\kappa_l$, $\Delta$) have been reset according to
ref \cite{FACCIOLI98}.

 The corresponding result for $R_{TL}(x_1,x_2,Q^2)$ is shown also in
Fig. 3.  We can see that the ratio obtained from a fully
non-relativistic model, lies reasonably close to the curve obtained
employing a relativistic kinetic energy but neglecting Melosh
rotations. Hence, we can safely conclude that $R_{TL}(x_1,x_2,Q^2)$ is
rather insensitive to the details of the mass operator, but very
sensitive to the relativistic spin corrections derived from a
light-front dynamics. This is not so evident for the values of $R_{T}$
or $R_{L}$ which are roughly doubled by the same change in the mass
operator (see Fig. 2).  This fact gives support to the suitability of
the chosen observable.  Moreover, due to this insensitivity to the
details of the spatial wave function, we do not expect a dramatic
change of our conclusions for other more sophisticated potential
models \cite{CAPSTICK86}.

\subsection{Experimental detection of relativistic spin effects}

	In order to study the feasibility of the experimental
detection of the differences for $R_{TL}$ we have to estimate the
expected errors for $R_{TL}$ under the conditions of RHIC and
HERA--$\vec{N}$. We have calculated the statistical errors according
to the expression \cite{MARTIN98}:

\begin{equation}
\delta A_{TT} = \frac{1}{P_B P_T \sqrt{{\cal L} \int \epsilon
d\sigma}}
\end{equation} 

\noindent
where $P_B$ ($P_T$) is the degree of polarization of the beam
(target), ${\cal L}$ is the expected luminosity and $\epsilon$ is the
efficiency in the detection of events. The unpolarized cross section
$d \sigma$ is integrated over bins of $Q$ and $y$.

	For RHIC we have taken \cite{MARTIN98} $P_B = P_T = 0.7$ and a
luminosity of 240 pb$^{-1}$. The calculated errors for the figure 3 in
the central rapidity region largely exceeds the separation between the
two main curves, even by assuming a 100 \% efficiency and integrating
over an interval $\Delta y=3$ and $\Delta Q =2$ around the central
values. Therefore, these effects will be unlikely to be observed at
RHIC.

For HERA--$\vec{N}$ the expected degrees of polarization are
\cite{SAITO98} $P_B=0.6$ and $P_T=0.8$ with a projected luminosity of
${\cal L} = 240$ pb$^{-1}$. The center of mass energy will be
$\sqrt{s}=39.2$ GeV, corresponding to a $E_{Beam} = 820$ GeV.  We have
checked that, though the error bars are far smaller than those
obtained for RHIC, relativistic spin effects cannot be disentangled
for $R_{TL}$ in the region $Q=3$ GeV, $y=0$.

	In order to maximize the rates we will integrate over the
whole range of $y$:

\begin{equation}
R_{TL}(Q^2)= \frac{\int (\sum_a e_a^2 h_1^a(x_1,Q^2)
h_1^{\bar{a}}(x_2,Q^2) + (x_1 \leftrightarrow x_2)) dy } {\int (\sum_a
e_a^2 g_1^a(x_1,Q^2) g_1^{\bar{a}}(x_2,Q^2) + (x_1 \leftrightarrow
x_2)) dy } \; .
\label{ryintdef}
\end{equation}  

The results for this ratio for the kinematics of HERA--$\vec{N}$ are
shown in figure 5, where we can observe the same pattern of
differences as the one seen in the y-dependent ratio (Fig. 3). The
relative insensitivity to the details of the chosen potential is also
evident in this representation. In the error bars shown in Fig. 5 we
have taken into account the limited acceptance of the detectors, as
explained in \cite{MARTIN98}, which roughly gives $\epsilon = 0.5-0.6$
in the considered region. We have also assumed that $\delta \epsilon =
\epsilon$. A measurement in the region $Q > 5$ GeV can not single out
which is the right spin-flavor basis, while experiments in the
low mass region ($Q \approx 3$ GeV) could be more selective, 
though some overlap between the error bars
still persists. For RHIC the acceptance corrections ($\epsilon \approx
0.1-0.2$ ) are too large to reveal relativistic differences.

\section{Conclusions and final remarks}
\label{CONCL}

        We have shown that Drell-Yan processes can 
probe the relativistic effects embodied in a  
light-front dynamical description of the low
energy spin structure of the nucleon. For this purpose, the ratio
$R_{TL}$ turns out to be particularly suitable since it is largely
independent of the structure of the mass operator and,
therefore, of the choice of the phenomenological interaction.

This insensitivity to the details of the spatial degrees of freedom is
not so evident for each asymmetry $R_{T}$ and $R_{L}$
separately. While predictions for $R_{TL}$ are robust, the given
values for $R_{T}$ and $R_{L}$ should be considered just at a
semiquantitative level. In fact, we have seen that relativistic spin
effects would reduce $R_{T}$ and therefore the chance of measure it
with respect to the 'maximal' scenario presented in
\cite{MARTIN98}, but no
definite quantitative conclusions about the feasibility of
measuring transversity should be drawn from Fig. 2.

By estimating the expected statistical errors for RHIC and HERA we
have concluded that for HERA--$\vec{N}$ it would be possible to
determine the right spin-flavour basis for the nucleon wave function
in the low mass region while this kind of measurement would be far
more difficult at RHIC.

        In the present work we have limited ourselves to give
predictions for D-Y related observables in some kinematic regions
accessible to RHIC, but the model can be straightforwardly applied to
other experimental conditions and to other kind of processes as
well. In particular, the observed enhancement of the transverse
polarization with respect to the longitudinal one may have some impact
in the extraction of the twist-3 contribution to the spin structure
function $g_2$ from DIS experiments \cite{G2EXP} (see also
\cite{CORTES92}).

        Finally let us emphasize that we have used a valence quark model
to evaluate the parton distributions at the hadronic scale, neglecting
sea quark (and antiquark) distributions at that scale. One could argue
that the consideration of a non-vanishing $h_1^{\bar{a}}(x,Q^2_0)$ and
$g_1^{\bar{a}}(x,Q^2_0)$ would mask the effects due to Melosh
rotations. Though a quantitative study of the change in the initial
conditions would require more elaborate models (beyond the valence
picture \cite{PREP}) there are, however, some qualitative reasons to
support the stability of the results presented in Figs. 3 and 5.  A
non-vanishing $h_1^{\bar{a}}(x,Q^2_0)$ (and $g_1^{\bar{a}}(x,Q^2_0)$)
has two main consequences on the overall scheme: firstly it slightly
raises the initial scale $Q^2_0$ and on the other hand it changes the
small $x$ behavior of the parton distributions. The (small) increment
of the initial scale would produce a little shift in the $Q^2$ scale
in Fig. 5, but not large enough to give rise to a confusion between
the two main curves. With respect to the change of the small $x$
behavior, it should be noticed that in the central rapidity regions we
are exploring ranges of $x$ around $0.1$ and also for the y-integrated 
observables this central region is dominant. 
Furthermore, since we deal with ratios we expect
the influence of these changes to be minimized.
\vskip 0.5cm

\section*{Acknowledgments}

We are grateful to S. Scopetta for useful suggestions and for
providing us with an updated version of the pQCD evolution code. We
thank V. Vento for his encouragement and help during the present study
and for a critical reading of the manuscript. We also thank O. Martin
for useful correspondence about the calculation of the statistical
errors and M. Miyama for sending us his evolution code for $h_1$
\cite{HIRAI98}.



\newpage

\begin{center}
{\bf Figure captions}
\end{center}

\begin{description}

\item{\bf Figure 1} Helicity and transversity distributions for the
$u$ quark (a) at the hadronic scale $Q_0^2=0.094$ GeV$^2$ and (b)
after evolution up to $Q^2=100$ GeV$^2$. In fig. a) the solid line
corresponds to $ x h_1$, the dashed line to $ x g_1$ and the dotted
line is the result when Melosh rotation is not considered
($h_1=g_1$). In Fig. (b) the solid and dashed lines represent $h_1$
and $g_1$ respectively. The dotted and dash-dotted lines correspond to
$h_1$ and $g_1$ when Melosh Rotation is neglected.

\item{\bf Figure 2.} $R_L$ (a) and $R_T$ (b) ratios as defined in
Eqs. (\ref{rldefinition},\ref{rtdefinition}) as a function of the
center of mass rapidity $y$ for $Q^2=100$ GeV$^2$ and a center of mass
energy $\sqrt{s} = 100$ GeV (solid line). The dashed line corresponds
to the case when the Melosh rotation is switched off.  The dotted line
is the result obtained in the non-relativistic model discussed in the
text.

\item{\bf Figure 3.} Ratio between transverse and longitudinal parton
distributions (eq. (\ref{rdef})) as a function of the center of mass
rapidity $y$ for $Q^2=100$ GeV$^2$ and a center of mass energy
$\sqrt{s} = 100$ GeV. Notation as in Fig. 2.

\item{\bf Figure 4.} Contributions to $\frac{\partial
g_1^{\bar{u}}}{\partial \ln Q^2}$ at LO coming from the terms
proportional to the gluons (solid line) and to the quarks + antiquarks
(dashed line). All the results correspond to $Q^2= 100$ GeV$^2$.

\item{\bf Figure 5.} Ratio between transverse and longitudinal parton
distributions, Eq. (\ref{ryintdef}), as a function of the invariant
mass of the produced lepton pair ($Q^2$) at a center of mass energy
corresponding to HERA--$\vec{N}$ ($\sqrt{s} = 39.2$) GeV. Notation as
in Fig. 2. Error bars have been calculated at LO and include
acceptance corrections. Error bars in the lower curve have been
slightly shifted to appreciate the overlap.

\end{description}

\newpage
\ 
\vfill
\begin{tabular}{lr}
\begin{tabular}{c}
\psfig{file=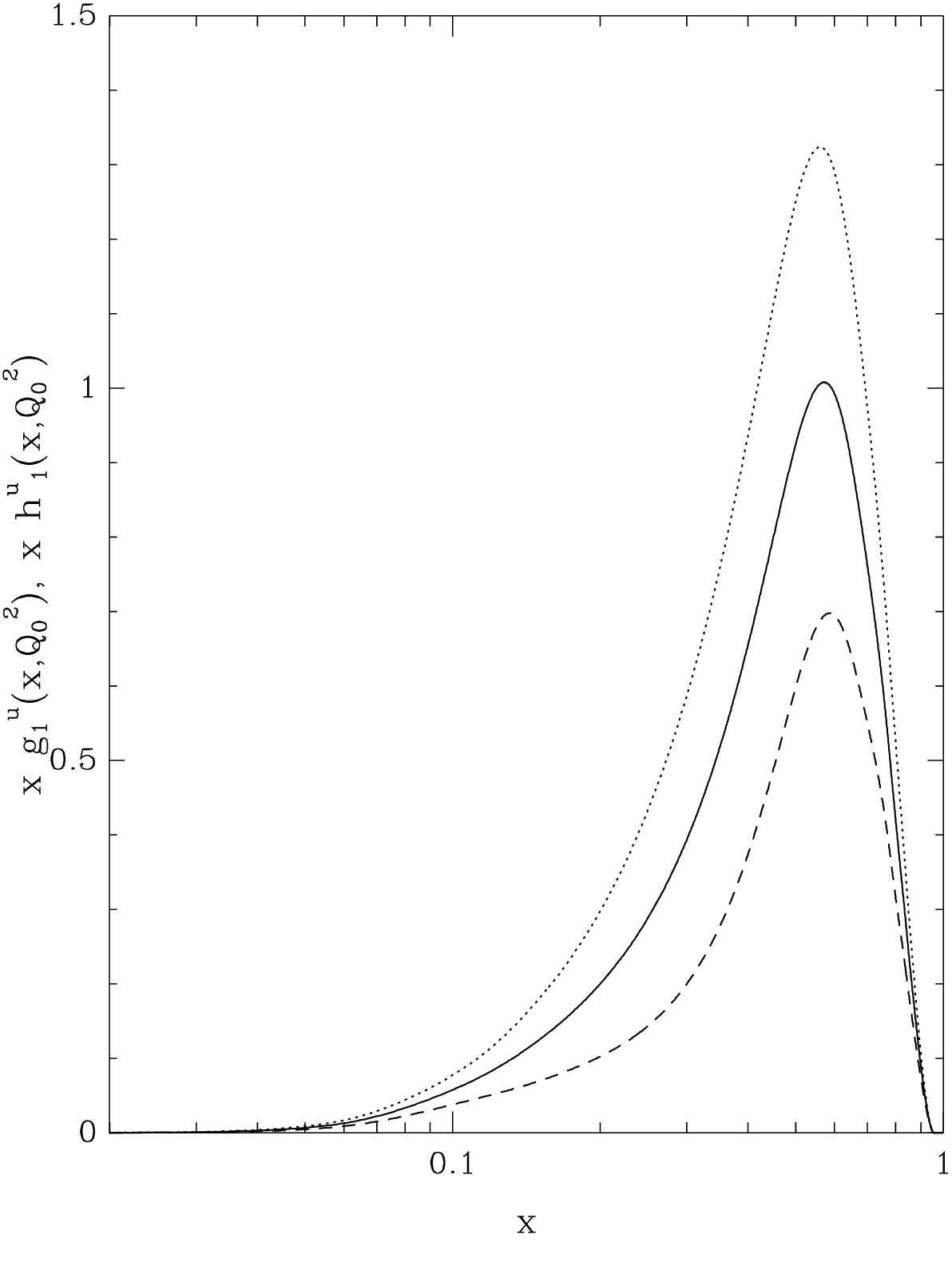,width=0.45\textwidth} \rule{1.ex}{0pt}  \\ \rule{0pt}{4.ex}{\bf
(a)}
\end{tabular} & 
\begin{tabular}{c}
\rule{1.ex}{0pt} \psfig{file=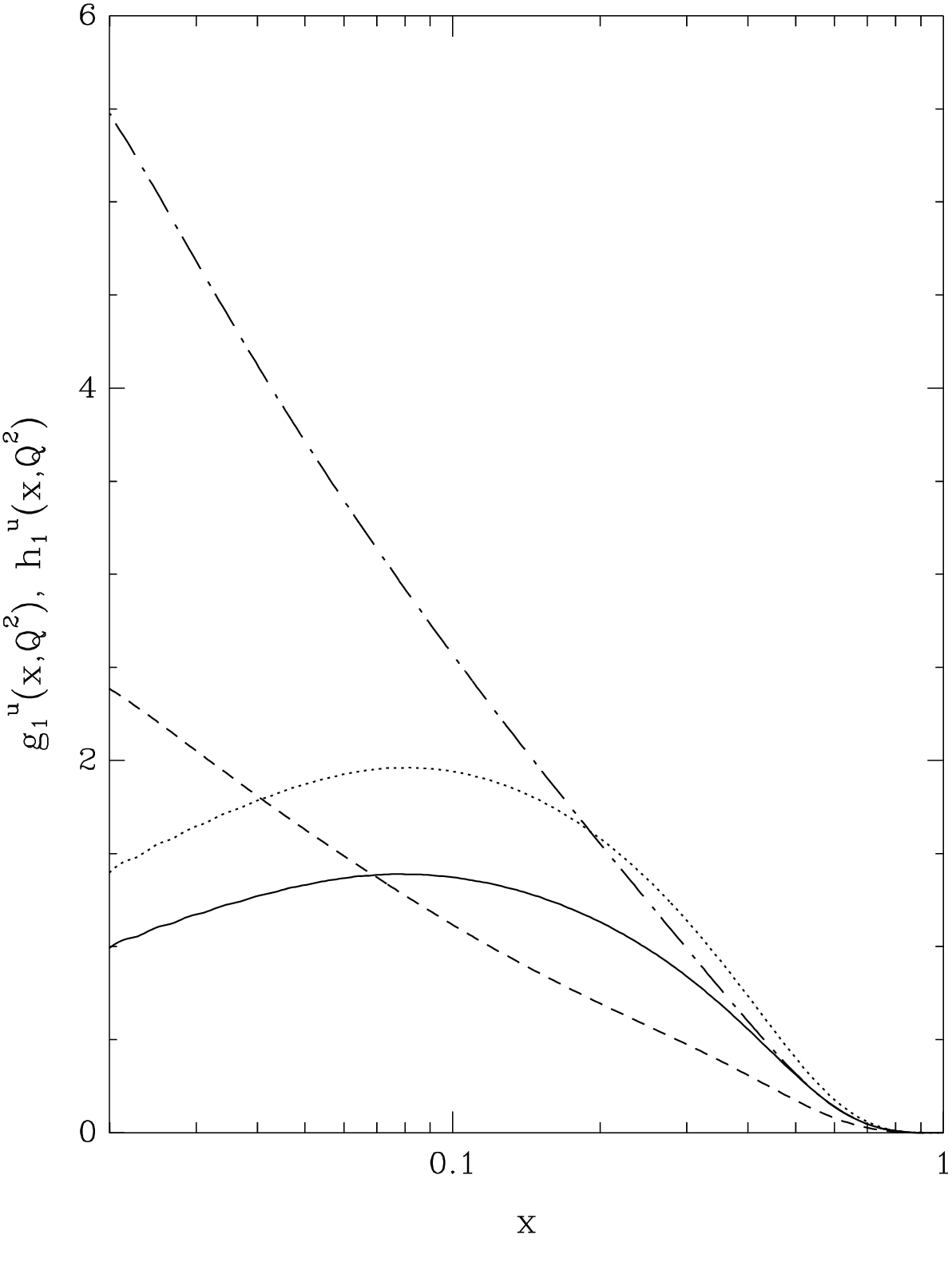,width=0.45\textwidth} \\ \rule{0pt}{4.ex}{\bf
(b)}
\end{tabular} 
\end{tabular}
\vskip 1cm 
\centerline{\bf Figure 1} 
\vfill
\newpage
\ 
\vfill
\begin{tabular}{lr}
\begin{tabular}{c}
\psfig{file=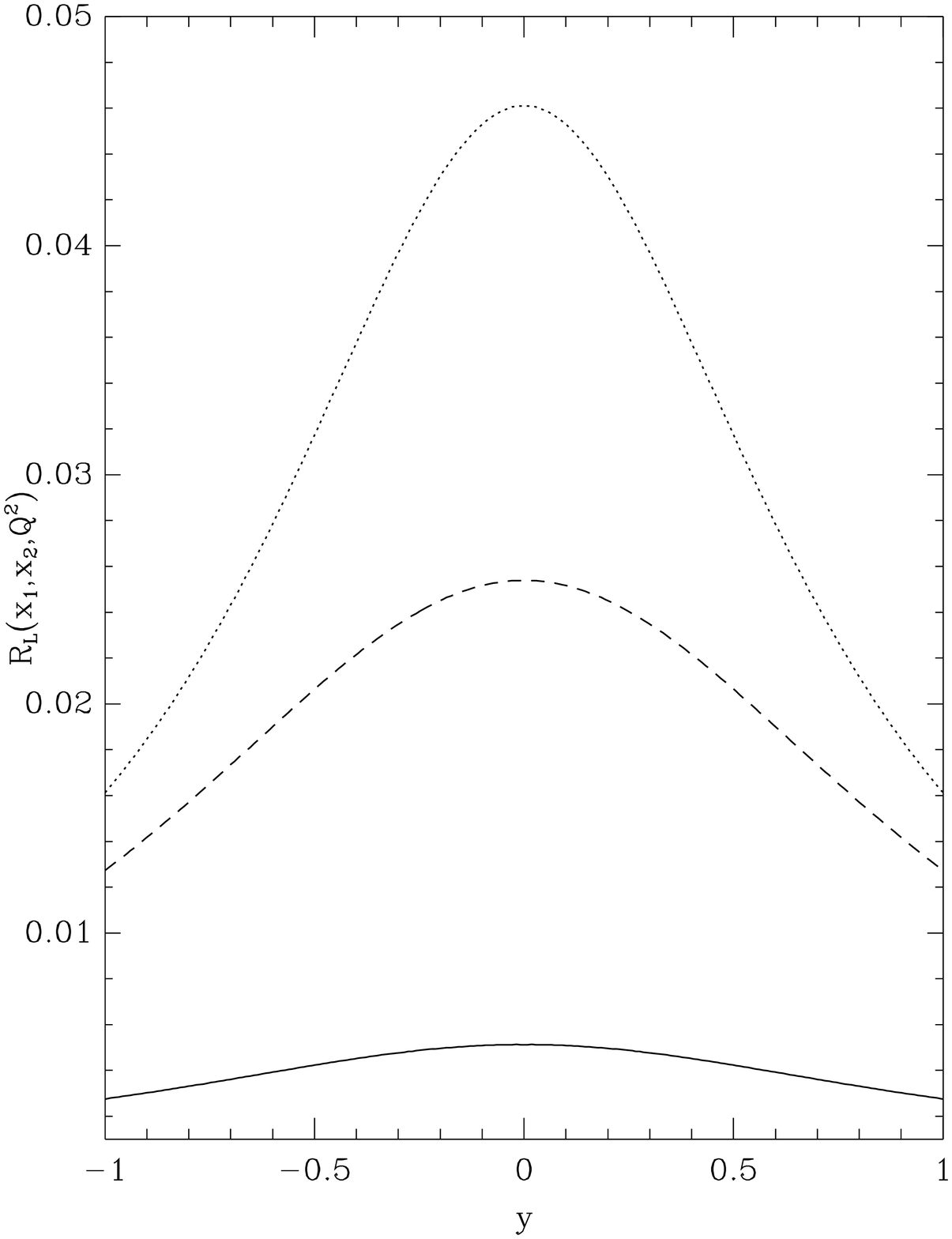,width=0.45\textwidth} \rule{1.ex}{0pt} \\ \rule{0pt}{4.ex}{\bf (a)}
\end{tabular} & 
\begin{tabular}{c}
\rule{1.ex}{0pt} \psfig{file=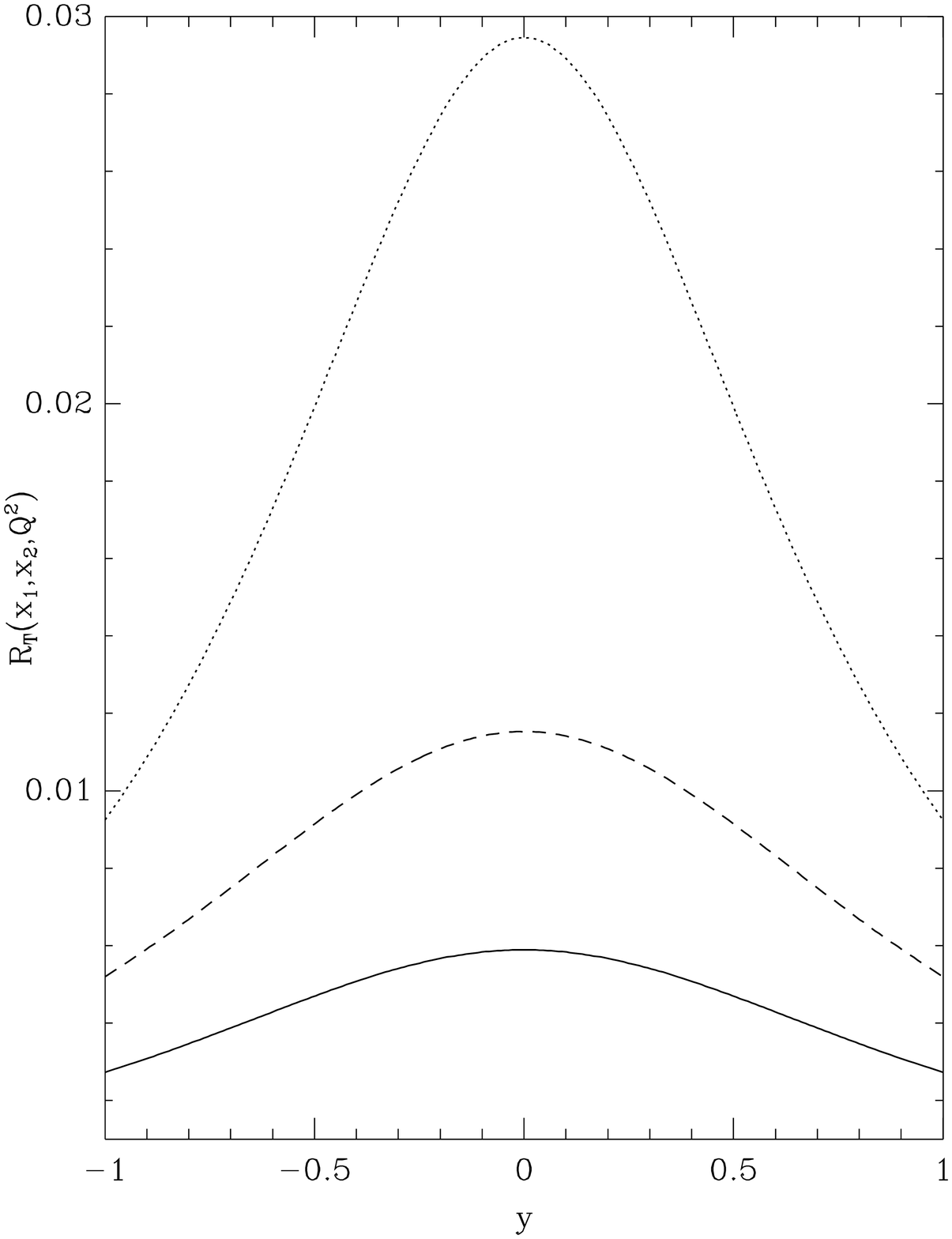,width=0.45\textwidth} \\ \rule{0pt}{4.ex}{\bf (b)}
\end{tabular} 
\end{tabular}
\vskip 1cm 
\centerline{\bf Figure 2} 
\vfill
\newpage
\ 
\vfill
\centerline{\protect\hbox{\psfig{file=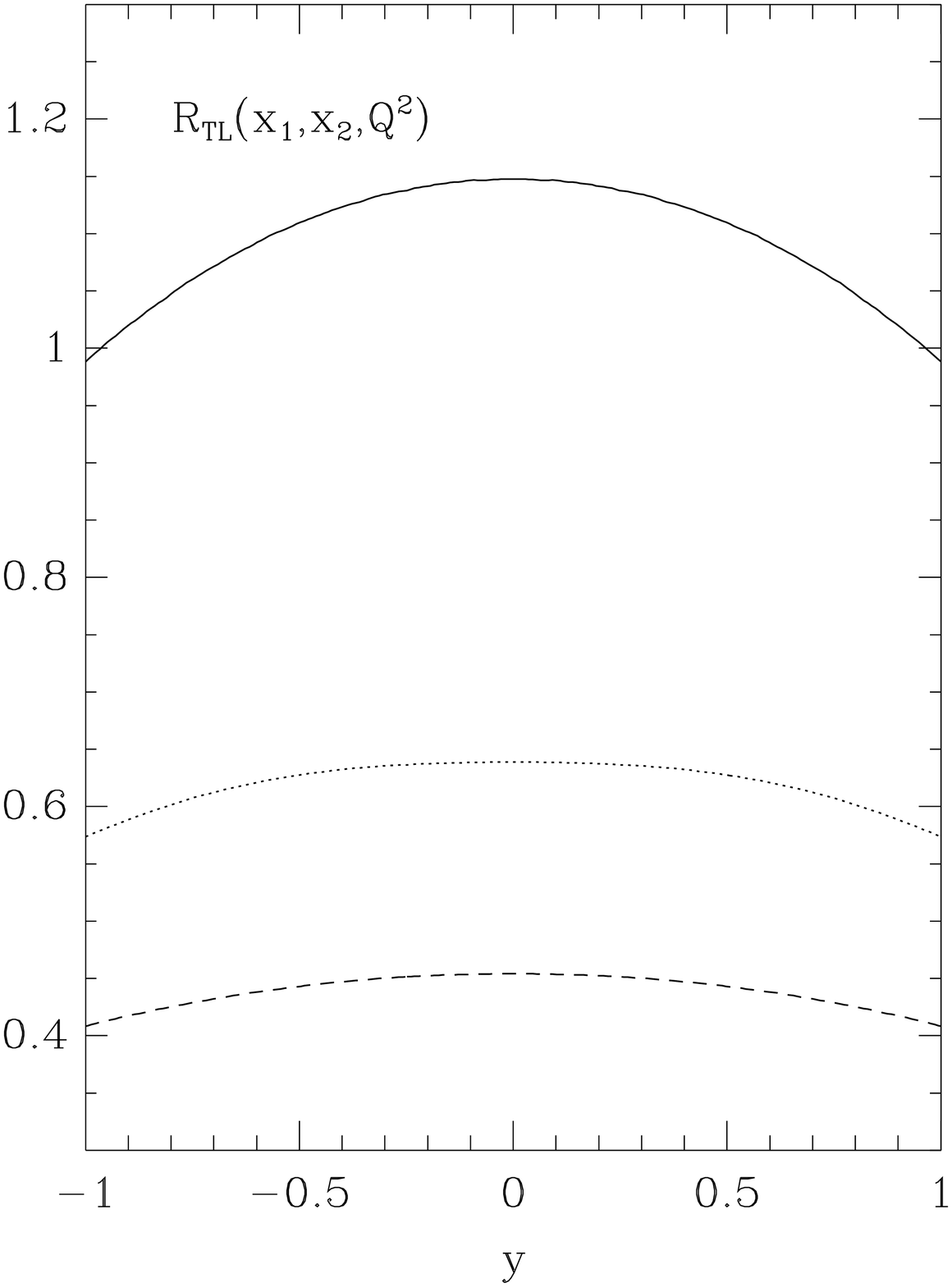,width=0.5\textwidth}}}
\vskip 1cm 
\centerline{\bf Figure 3} 
\vfill
\newpage
\ 
\vfill
\centerline{\protect\hbox{\psfig{file=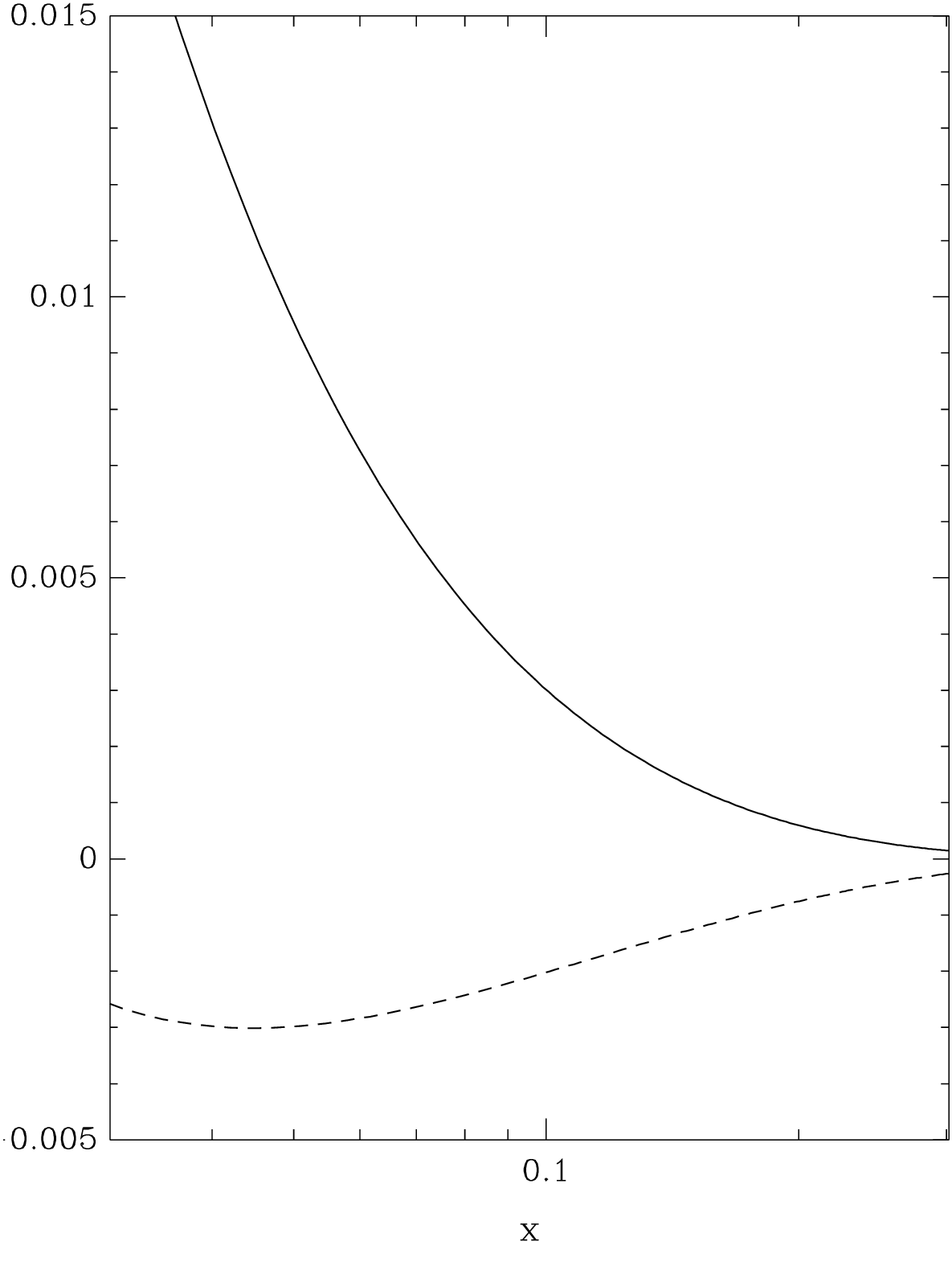,width=0.5\textwidth}}}
\vskip 1cm 
\centerline{\bf Figure 4} 
\vfill
\newpage
\ 
\vfill
\centerline{\protect\hbox{\psfig{file=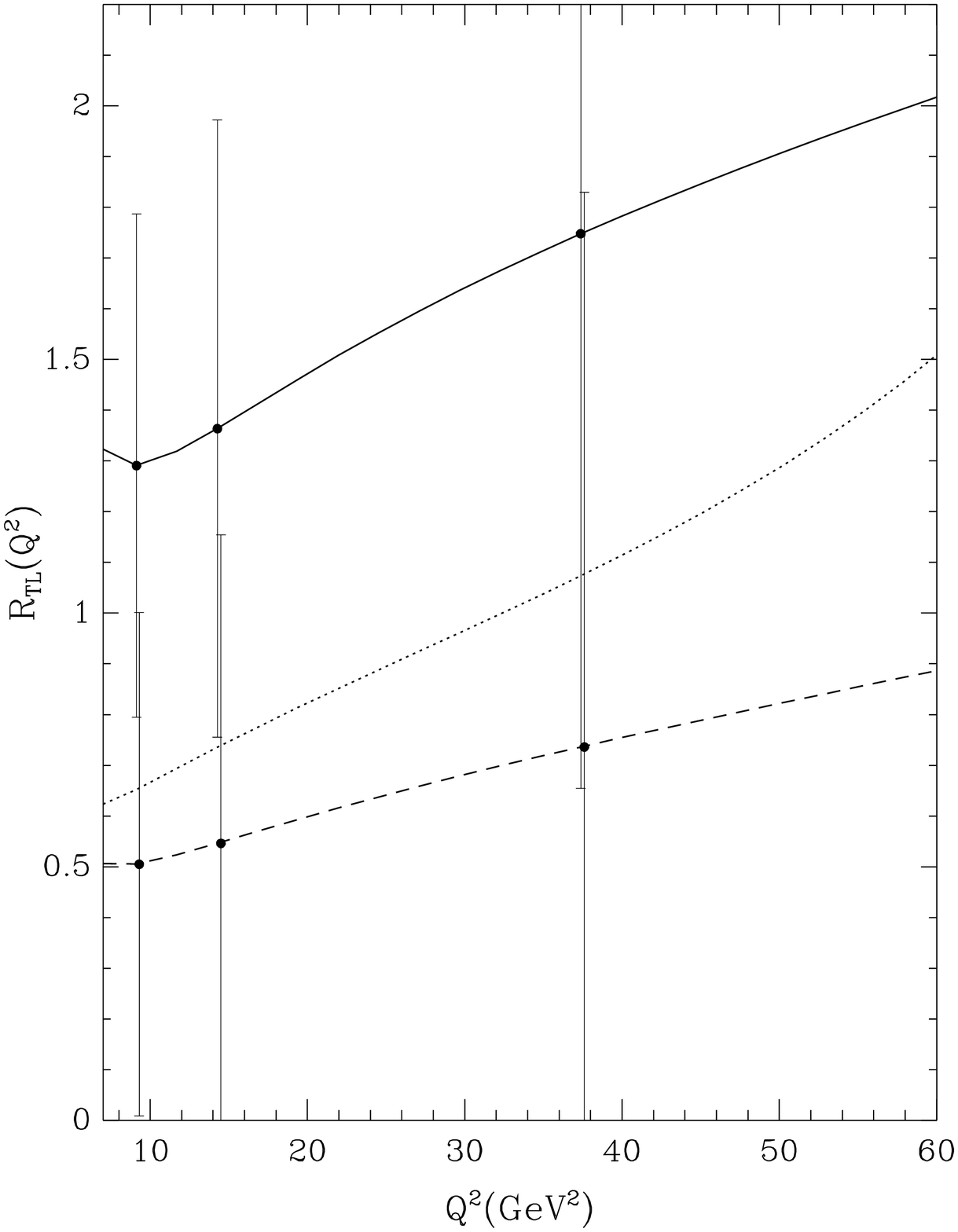,width=0.5\textwidth}}}
\vskip 1cm 
\centerline{\bf Figure 5} 
\vfill

\end{document}